\documentclass{article}
\usepackage{amssymb}
\usepackage{amsfonts}
\usepackage{amsmath}

\input{tcilatex}
\begin{document}

\title{\textbf{Black Holes,}\\
\textbf{Gravitational Waves}\\
\textbf{and Quantum Gravity}}
\author{W. F. Chagas-Filho \\
Physics Department, Federal University of Sergipe, Brazil\\
e-mail: wfilho@ufs.br}
\maketitle

\begin{abstract}
Loop Quantum Gravity is a theory that attempts to describe the quantum
mechanics of the gravitational field based on the canonical quantization of
General Relativity. According to Loop Quantum Gravity, in a gravitational
field, geometric quantities such as area and volume are quantized in terms
of the Planck length. In this paper we present the basic ideas for a future,
mathematically more rigorous, attempt to combine black holes and
gravitational waves using the quantization of geometric quantities
introduced by Loop Quantum Gravity.
\end{abstract}

\section{Introduction}

A black hole is a region of the universe where the causal structure of the
spacetime is so deformed by gravity that even light can not scape from this
region. The existence of black holes in the universe is one of the
implications of Einstein%
\'{}%
s field equations for General Relativity (GR). Today, indirect as well as
direct evidence has accumulated confirming the existence of black holes. The
simplest black hole occurs in the Schwarzschild spacetime. The Schwarzschild
spacetime is an exact, spherically symmetric, solution of the Einstein
equations describing the gravitational field in the exterior region of a
mass M with no electric charge and no angular momentum. The geometric
structure of the Schwarzschild spacetime is characterized by the existence
of a particular value $R_{S}$ for the radial coordinate below which it
becomes impossible to scape form the gravitational attraction of the black
hole. The spherical surface defined by $R_{S}$ is called the event horizon
of the black hole.

In 1975 Stephen Hawking [1], using the methods of quantum field theory in
curved spacetime, showed that black holes are not really black because, as a
consequence of quantum vacuum fluctuations near the event horizon, a black
hole can emit quantum particles in the thermal spectrum. As a consequence of
this thermal radiation a black hole has a temperature and an associated
entropy and can ultimately evaporate. A fundamental open conceptual problem
associated with the Hawking thermal radiation is the so-called information
loss paradox. Energy is carried away by the Hawking radiation, so that the
black hole eventually evaporates away entirely, leaving a future with the
causal structure of Minkowski space. Information that falls past the event
horizon, for instance the black hole mass, appears to be lost. For a review
see refs. [2,3,4,5,6]. In the literature a black hole with Hawking radiation
is termed a \textsl{semi-classical black hole}, to distinguish it from the
classical black hole described by GR. The Hawking radiation is a tiny effect
for most black holes. For example, a black hole with 15 times the mass of
our Sun has a temperature of $4,1\times 10^{-9}K$ and the time for the black
hole to evaporate all of its mass by means of Hawking radiation is given by $%
2,2\times 10^{78}s$ which is about $60$ orders of magnitude larger than the
age of the Universe [7].

In quantum field theory in curved spacetime, one treats gravitation
classically, as in the framework of GR. Thus, spacetime structure is
described by a manifold $M$, on which is defined a classical, Lorentz
signature metric $g_{\mu \nu }$. One thereby avoids confronting the
fundamental difficulty of how to formulate quantum field theory without a
classical background metrical (and causal) structure of spacetime. One
expects that quantum field theory in curved spacetime should have only a
limited range of validity [4]. In particular, it certainly should break
down, and be replaced by a quantum theory of gravitation coupled to matter,
when the spacetime curvature approaches Planck scales [4]. In this paper we
suggest the possible existence of another mechanism for the emission of
quantum radiation by a black hole. Here we are interested in a mechanism of
emission of gravitational radiation by a \textsl{quantum black hole}. The
basic concept that supports the ideas presented in this paper is that of the
Planck scale. Let us briefly review the Planck scale.

In 1899 Max Planck [8] noticed that by combining three of the fundamental
constants of physics, the Newtonian gravitational constant $G$, the speed of
light $c$ and Planck%
\'{}%
s constant 
h{\hskip-.2em}\llap{\protect\rule[1.1ex]{.325em}{.1ex}}{\hskip.2em}
in a unique way, he could define a fundamental scale of length, time and
mass. Today this fundamental scale is called the Planck scale. It is given by

1) the Planck length 
\begin{equation}
L_{P}=\sqrt{\frac{\text{%
h{\hskip-.2em}\llap{\protect\rule[1.1ex]{.325em}{.1ex}}{\hskip.2em}%
G}}{c^{3}}}=1,62\times 10^{-35}m  \tag{1}
\end{equation}

2) the Planck time 
\begin{equation}
T_{P}=\sqrt{\frac{\text{%
h{\hskip-.2em}\llap{\protect\rule[1.1ex]{.325em}{.1ex}}{\hskip.2em}%
G}}{c^{5}}}=5,40\times 10^{-44}s  \tag{2}
\end{equation}

3) the Planck mass 
\begin{equation}
M_{P}=\sqrt{\frac{\text{%
h{\hskip-.2em}\llap{\protect\rule[1.1ex]{.325em}{.1ex}}{\hskip.2em}%
c}}{G}}=2,17\times 10^{-5}g  \tag{3}
\end{equation}%
In this paper we will adopt a particular interpretation of the Planck scale.
This particular interpretation can be justified as follows. First, observe
that the Planck length is the distance that light travels during the Planck
time. Therefore the existence of the Planck length $L_{P},$ together with
the constancy of the speed of light $c,$ automatically defines the Planck
time $T_{P}=L_{P}/c.$ Observe also that the Planck mass $M_{P}$ can be
written in terms of $L_{P}$ as 
\begin{equation}
M_{P}=\frac{c^{2}}{G}L_{P}  \tag{4}
\end{equation}%
The above observations can be interpreted as indications that the Planck
length $L_{P}$ is the minimum length in our universe and that the Planck
time $T_{P}$ and the Planck mass $M_{P}$ can be obtained from this minimum
length using the fundamental constants $c$ and $G$. It is this particular
interpretation of the Planck scale that we will adopt in this paper.

The existence of gravitational waves was also one of the predictions of
Einstein field equations for GR. Gravitational waves were finally detected
in 2015. Contrary to black holes, which are associated with strong
gravitational fields, gravitational waves are usually associated with weak
gravitational fields. The most direct way to describe the propagation of
gravitational waves is to expand the curved spacetime metric $g_{\mu \nu }$
around the flat spacetime metric $\eta _{\mu \nu }$ and retain only the
linear order terms in the expansion. This procedure leads to the linearized
Einstein equations. The difficulty of this approach is to separate the
physical from the unphysical degrees of freedom described by the
gravitational wave. In this paper we describe in detail the procedure for
removing the unphysical degrees of freedom contained in the metric that
describes a gravitational wave. This is a necessary step to give further
support to the quantization of the energy of a gravitational wave described
in [9]. It will also support the construction of a quantum gravitational
wave described in this paper. Here we will explain how to relate black holes
to gravitational waves using the quantization of geometric quantities
discovered in the quantum theory of gravity called Loop Quantum Gravity
(LQG).

LQG [10,11,12] is a theory that attempts to describe the quantum mechanics
of the gravitational field based on the canonical quantization of GR. The
construction of LQG only became possible after 1986, when Ashtekar [13]
introduced a new set of canonical variables for describing GR. Instead of
the traditional metric tensor field $g_{\mu \nu }$ of GR as the
configuration variable, Ashtekar introduced an $SU(2)$ connection $A_{\mu
}^{i}$ as the configuration variable. This allowed the description of GR as
a constrained Hamiltonian system with first-class constraints [14] only. The
most striking result of LQG is that, in a gravitational field, geometric
quantities such as area and volume are quantized in terms of the Planck
length $L_{P}$ given in equation (1).

The paper is organized as follows. In section two we review the Shwarzschild
solution to the field equations of GR and the basic equations of black hole
Thermodynamics. In section three we review in detail the description of the
propagation of gravitational waves in the transverse-traceless gauge. In
section four we present our contribution to these subjects. We explain how
we can construct a quantum gravitational wave and a quantum equation for the
Schwarzschild black hole entropy. Finally we explain how a quantum black
hole can convert all of its mass into quantum gravitational radiation. We
present our conclusions in section five.

\section{The Schwarzschild black hole}

Einstein equations can be derived from the Einstein-Hilbert action%
\begin{equation}
S=\frac{c^{3}}{16\pi G}\int d^{4}x\sqrt{\left[ -\det g_{\mu \nu }\right] }%
R+S_{M}  \tag{5}
\end{equation}%
where $S_{M}$ is the matter action. The energy-momentum tensor of matter, $%
T^{\mu \nu }$, is defined from the variation of the matter action under a
change of the spacetime metric $g_{\mu \nu }\rightarrow g_{\mu \nu }+\delta
g_{\mu \nu }$ according to 
\begin{equation}
\delta S_{M}=\frac{1}{2c}\int d^{4}x\sqrt{\left[ -\det g_{\mu \nu }\right] }%
T^{\mu \nu }\delta g_{\mu \nu }  \tag{6}
\end{equation}%
We must now define the relevant geometric quantities obtained from the
spacetime metric $g_{\mu \nu }$. The first of these are the Christoffel
symbols%
\begin{equation}
\Gamma _{\mu \nu }^{\rho }=\frac{1}{2}g^{\rho \sigma }\left( \partial _{\mu
}g_{\sigma \nu }+\partial _{\nu }g_{\sigma \mu }-\partial _{\sigma }g_{\mu
\nu }\right)  \tag{7}
\end{equation}%
In this equation $g^{\rho \sigma }$ is the inverse metric of $g_{\rho \sigma
}$. From the Christoffel symbols we define the Riemann curvature tensor 
\begin{equation}
R_{\nu \rho \sigma }^{\mu }=\partial _{\rho }\Gamma _{\nu \sigma }^{\mu
}-\partial _{\sigma }\Gamma _{\nu \rho }^{\mu }+\Gamma _{\alpha \rho }^{\mu
}\Gamma _{\nu \sigma }^{\alpha }-\Gamma _{\alpha \sigma }^{\mu }\Gamma _{\nu
\rho }^{\alpha }  \tag{8}
\end{equation}%
Contracting the Riemann tensor we obtain the Ricci tensor 
\begin{equation}
R_{\mu \nu }=R_{\mu \alpha \nu }^{\alpha }  \tag{9}
\end{equation}%
Contracting again we obtain the Ricci scalar or scalar curvature 
\begin{equation}
R=g^{\mu \nu }R_{\mu \nu }  \tag{10}
\end{equation}%
Varying the Einstein-Hilbert action (5) with respect to $g_{\mu \nu }$ we
obtain the field equations of GR 
\begin{equation}
R_{\mu \nu }-\frac{1}{2}g_{\mu \nu }R=\frac{8\pi G}{c^{4}}T_{\mu \nu } 
\tag{11}
\end{equation}
$T_{\mu \nu }$ describes the flow of energy and momentum through a given
point in spacetime.

As mentioned in the introduction, the Schwarzschild spacetime describes the
gravitational field in the outside region of a mass $M$ with no electric
charge and no angular momentum. In this case the Einstein equations become
the vacuum equations 
\begin{equation}
R_{\mu \nu }=0  \tag{12}
\end{equation}%
with the Schwarzschild metric 
\begin{equation}
ds^{2}=-\left( 1-\frac{2GM}{c^{2}r}\right) dt^{2}+\left( 1-\frac{2GM}{c^{2}r}%
\right) ^{-1}dr^{2}+r^{2}(d\theta ^{2}+\sin ^{2}\theta d\varphi ^{2}) 
\tag{13}
\end{equation}%
as the only exact solution. Notice that the first term on the right hand
side vanishes and the second becomes divergent when $r=R_{S}=2GM/c^{2}$.
This value for $r$ is called the Schwarzschild radius and the spherical
surface associated with it is called the event horizon for the Schwarzschild
black hole. The event horizon acts as a one-way membrane: matter and energy
can go in but, once inside, can never go out. For details see [7,15].

The mass $M$ of the central gravitating body is related to the horizon area $%
A$ by 
\begin{equation}
M=\sqrt{\frac{c^{3}A}{16\pi G^{2}}}  \tag{14}
\end{equation}%
Advanced methods in field theory that consider quantum vacuum fluctuations
that occur in the vicinity of the event horizon find the result that the
Schwarzschild black hole can emit quantum particles in the thermal spectrum.
This thermal emission is called the Hawking radiation. The distribution of
energies emitted by the black hole as Hawking radiation is equivalent to
that of a blackbody with a temperature proportional to the surface gravity
of the black hole. Specifically, the black hole temperature is given by 
\begin{equation}
T=\frac{\text{%
h{\hskip-.2em}\llap{\protect\rule[1.1ex]{.325em}{.1ex}}{\hskip.2em}%
c}^{3}}{8\pi k_{B}GM}  \tag{15}
\end{equation}%
where $k_{B}$ is the Boltzman constant. The temperature (15) is proportional
to 
h{\hskip-.2em}\llap{\protect\rule[1.1ex]{.325em}{.1ex}}{\hskip.2em}
so it is a quantum effect that vanishes in the 
h{\hskip-.2em}\llap{\protect\rule[1.1ex]{.325em}{.1ex}}{\hskip.2em}%
$\rightarrow 0$ classical limit. As a consequence of the temperature (15)
the black hole has an entropy given by 
\begin{equation}
S=\frac{c^{3}k_{B}}{4\text{%
h{\hskip-.2em}\llap{\protect\rule[1.1ex]{.325em}{.1ex}}{\hskip.2em}%
G}}A=\frac{k_{B}A}{4L_{P}^{2}}  \tag{16}
\end{equation}

We now briefly describe how equation (16) for the black hole entropy emerges
in the framework of quantum gravity. In LQG the area of a surface in a
gravitational field is quantized. The area $A$ of the event horizon of a
black hole is given by [10] 
\begin{equation*}
A=8\pi \gamma \frac{\text{%
h{\hskip-.2em}\llap{\protect\rule[1.1ex]{.325em}{.1ex}}{\hskip.2em}%
G}}{c^{3}}\Sigma _{i}\sqrt{j_{i}(j_{i}+1)}
\end{equation*}%
\begin{equation}
=8\pi \gamma L_{P}^{2}\Sigma _{i}\sqrt{j_{i}(j_{i}+1)}  \tag{17}
\end{equation}%
where $\gamma $ is the Immirzi parameter [16], used to fix the exact scale
of the quantum theory, and $j_{i}=j_{1},...,j_{n}$ are the spins of the
links intersecting the event horizon surface. The black hole entropy is
given by%
\begin{equation}
S=k_{B}\ln N(A)  \tag{18}
\end{equation}%
where $N(A)$ is the number of states that the geometry of a surface with
area $A$ can have. The possible states are obtained by considering all sets
of $j_{i}$ that give the area $A$ and, for each set, the dimension of $%
\otimes _{i}\QTR{sl}{H}_{i}$ where $H_{i}$ is the representation space of
the spin $j_{i}$.

In LQG it was first assumed [17] that the number of possible states is
dominated by the case $j=1/2$. In this case the quantum of area is given by%
\begin{equation}
A_{\frac{1}{2}}=4\pi \sqrt{3}\gamma L_{P}^{2}  \tag{19}
\end{equation}%
Hence there are 
\begin{equation}
n=\frac{A}{A_{\frac{1}{2}}}=\frac{A}{4\pi \sqrt{3}\gamma L_{P}^{2}}  \tag{20}
\end{equation}%
intersections and the dimension of $H_{1/2}=2\frac{1}{2}+1=2$. So the number
of quantum states of the event horizon area $A$ is 
\begin{equation}
N(A)=2^{n}=2^{A/4\pi \sqrt{3}\gamma L_{P}^{2}}  \tag{21}
\end{equation}%
and the black hole entropy is%
\begin{equation}
S=\frac{\ln 2}{4\pi \sqrt{3}\gamma }\frac{k_{B}A}{L_{P}^{2}}  \tag{22}
\end{equation}%
which agrees with equation (16) if we choose 
\begin{equation}
\gamma =\frac{\ln 2}{\pi \sqrt{3}}  \tag{23}
\end{equation}%
Later, it was realized [18,19] that the number of possible states could
instead be dominated by the case $j=1$ and a similar calculation was
performed for the spin $j=1$. Now the black hole entropy given by equation
(16) is again reproduced provided we choose the Immirzi parameter $\gamma $
to be 
\begin{equation}
\gamma =\frac{\ln 3}{2\pi \sqrt{2}}  \tag{24}
\end{equation}%
which in turn fixes the minimal quantum of area to be%
\begin{equation}
A_{1}=4(\ln 3)L_{P}^{2}  \tag{25}
\end{equation}%
The same calculation can be performed with higher spins $j_{i}=\frac{3}{2}%
,2,...$ and equation (16) for the black hole entropy will always be obtained
provided we choose the appropriate value for the Immirzi parameter $\gamma $.

The situation in LQG described above leaves no doubts about the validity of
equation (16) in describing the entropy of the Schwarzschild black hole. But
the true microstate responsible for the entropy has not been determined yet.
It can be the quantum of area $A_{\frac{1}{2}}$, or the quantum of area $%
A_{1}$, or any quantum of area we choose, provided we select the appropriate
value of the Immirzi parameter $\gamma $ that reproduces the black hole
entropy equation (16). As we will see below, equation (16) for the black
hole entropy may not be the end of the story. This is because there is one
important information about the Schwarzschild spacetime which was not used
in LQG to arrive at the entropy (16). This important information is the
spherical symmetry of the event horizon.

\section{Gravitational waves}

General Relativity is invariant under a huge symmetry group, the group of
all possible coordinate transformations 
\begin{equation}
x^{\mu }\rightarrow \acute{x}^{\mu }(x)  \tag{26}
\end{equation}%
where $\acute{x}^{\mu }$ is an arbitrary smooth function of $x^{\mu }$.
Under the transformation (26) the spacetime metric transforms as 
\begin{equation}
g_{\mu \nu }(x)\rightarrow g%
{\acute{}}%
_{\mu \nu }(\acute{x})=\frac{\partial x^{\rho }}{\partial \acute{x}^{\mu }}%
\frac{\partial x^{\sigma }}{\partial \acute{x}^{\nu }}g_{\rho \sigma }(x) 
\tag{27}
\end{equation}%
This symmetry is the gauge symmetry of GR.

As a first step toward the description of gravitational waves we must expand
Einstein equations around the flat spacetime metric 
\begin{equation}
g_{\mu \nu }=\eta _{\mu \nu }+h_{\mu \nu }\text{ \ \ \ }\mid h_{\mu \nu
}\mid \text{ }\ll 1  \tag{28}
\end{equation}%
and retain only terms to linear order in $h_{\mu \nu }$. The resulting
theory is called the linearized theory [20].

After choosing a frame where equation (28) holds, a residual gauge symmetry
remains. Consider the transformation of coordinates 
\begin{equation}
x^{\mu }\rightarrow \acute{x}^{\mu }=x^{\mu }+\xi ^{\mu }(x)  \tag{29}
\end{equation}%
where the derivatives $\mid \partial _{\mu }\xi _{\nu }\mid $ are of the
same order of smallness as $\mid h_{\mu \nu }\mid $. Using the
transformation law of the metric, equation (27), we find that the
transformation of $h_{\mu \nu }$, to lowest order, is 
\begin{equation}
h_{\mu \nu }(x)\rightarrow h%
{\acute{}}%
_{\mu \nu }(\acute{x})=h_{\mu \nu }(x)-\left( \partial _{\mu }\xi _{\nu
}+\partial _{\nu }\xi _{\mu }\right)  \tag{30}
\end{equation}%
If $\mid \partial _{\mu }\xi _{\nu }\mid $ are of the same order of
smallness as $\mid h_{\mu \nu }\mid $, the condition $\mid h_{\mu \nu }\mid $
$\ll $ 1 is preserved. Now we are ready to construct the linearized version
of Einstein equations.

To linear order in $h_{\mu \nu }$ the Christoffel symbols are given by 
\begin{equation}
\Gamma _{\mu \nu }^{\rho }=\frac{1}{2}\eta ^{\rho \lambda }\left( \partial
_{\mu }h_{\nu \lambda }+\partial _{\nu }h_{\lambda \mu }-\partial _{\lambda
}h_{\mu \nu }\right)  \tag{31}
\end{equation}%
Lowering an index for convenience the Riemann tensor becomes 
\begin{equation}
R_{\mu \nu \rho \sigma }=\frac{1}{2}\left( \partial _{\rho }\partial _{\nu
}h_{\mu \sigma }+\partial _{\sigma }\partial _{\mu }h_{\nu \rho }-\partial
_{\sigma }\partial _{\nu }h_{\mu \rho }-\partial _{\rho }\partial _{\mu
}h_{\nu \sigma }\right)  \tag{32}
\end{equation}%
The Ricci tensor is given by 
\begin{equation}
R_{\mu \nu }=\frac{1}{2}\left( \partial _{\sigma }\partial _{\nu }h_{\mu
}^{\sigma }+\partial _{\sigma }\partial _{\mu }h_{\nu }^{\sigma }-\partial
_{\mu }\partial _{\nu }h-\square h_{\mu \nu }\right)  \tag{33}
\end{equation}%
where we have defined the trace of the perturbation as $h=\eta ^{\mu \nu
}h_{\mu \nu }=h_{\mu }^{\mu }$ and $\square =-\frac{1}{c^{2}}\partial
_{t}^{2}+\partial _{x}^{2}+\partial _{y}^{2}+\partial _{z}^{2}$. And finally
the Ricci scalar is 
\begin{equation}
R=\partial _{\mu }\partial _{\nu }h^{\mu \nu }-\square h  \tag{34}
\end{equation}%
The linearized equations of motion are written more compactly defining $\bar{%
h}_{\mu \nu }=h_{\mu \nu }-\frac{1}{2}\eta _{\mu \nu }h$. It is a
straightforward algebra to compute the linearized Einstein tensor $G_{\mu
\nu }=R_{\mu \nu }-\frac{1}{2}g_{\mu \nu }R$ and we find that the
linearization of the Einstein equations (11) gives 
\begin{equation}
\square \bar{h}_{\mu \nu }+\eta _{\mu \nu }\partial ^{\rho }\partial
^{\sigma }\bar{h}_{\rho \sigma }-\partial ^{\rho }\partial _{\nu }\bar{h}%
_{\mu \rho }-\partial ^{\rho }\partial _{\mu }\bar{h}_{\nu \rho }=-\frac{%
16\pi G}{c^{4}}T_{\mu \nu }  \tag{35}
\end{equation}%
We can now use the residual gauge freedom (29) to choose the Lorentz gauge 
\textsl{\ }%
\begin{equation}
\partial ^{\nu }\bar{h}_{\mu \nu }=0  \tag{36}
\end{equation}%
In this gauge the last three terms on the left-hand side of equation (35)
vanish and we get a simple wave equation 
\begin{equation}
\square \bar{h}_{\mu \nu }=-\frac{16\pi G}{c^{4}}T_{\mu \nu }  \tag{37}
\end{equation}%
Observe that the Lorentz gauge gives four conditions, that reduce the 10
independent components of the symmetric $4\times 4$ matrix $h_{\mu \nu }$ to
six independent components. Equations (36) and (37) together imply for
consistency that 
\begin{equation}
\partial ^{\nu }T_{\mu \nu }=0  \tag{38}
\end{equation}%
which is the conservation of energy-momentum in the linearized theory.

Equation (37) is the basic result for computing the generation of
gravitational waves within the linearized theory. To study the propagation
of gravitational waves we are rather interested in this equation outside the
source, where $T_{\mu \nu }=0$, 
\begin{equation}
\square \bar{h}_{\mu \nu }=0  \tag{39}
\end{equation}%
Outside the source we can greatly simplify the form of the metric, observing
that equation (36) does not fix the gauge completely. To see this, using the
symmetry transformation (30), we can impose the Lorentz gauge (36) and we
observe that, in terms of $\bar{h}_{\mu \nu }$, equation (30) becomes 
\begin{equation}
\bar{h}_{\mu \nu }\rightarrow \bar{h}%
{\acute{}}%
_{\mu \nu }=\bar{h}_{\mu \nu }-\left( \partial _{\mu }\xi _{\nu }+\partial
_{\nu }\xi _{\mu }-\eta _{\mu \nu }\partial _{\rho }\xi ^{\rho }\right) 
\tag{40}
\end{equation}%
and therefore 
\begin{equation}
\partial ^{\nu }\bar{h}_{\mu \nu }\rightarrow \left( \partial ^{\nu }\bar{h}%
_{\mu \nu }\right) 
{\acute{}}%
=\partial ^{\nu }\bar{h}_{\mu \nu }-\square \xi _{\mu }  \tag{41}
\end{equation}%
Equation (41) means that the Lorentz gauge condition (36) does not remove
all the unphysical degrees of freedom. As we see from (41), we can perform a
further coordinate transformation $x^{\mu }\rightarrow x^{\mu }+\xi ^{\mu }$
with 
\begin{equation}
\square \xi ^{\mu }=0  \tag{42}
\end{equation}%
and the Lorentz gauge (36) is not spoiled.

If $\square \xi _{\mu }=0$ then also $\square \xi _{\mu \nu }=0$, where 
\begin{equation}
\xi _{\mu \nu }=\partial _{\mu }\xi _{\nu }+\partial _{\nu }\xi _{\mu }-\eta
_{\mu \nu }\partial _{\rho }\xi ^{\rho }  \tag{43}
\end{equation}%
since the flat d`Alembertian $\square $ commutes with $\partial _{\mu }$.
Then equation (40) tells us that, from the six independent components of $%
\bar{h}_{\mu \nu }$, which satisfy $\square \bar{h}_{\mu \nu }=0$, we can
subtract the functions $\xi _{\mu \nu }$, which depend on four independent
arbitrary functions $\xi _{\mu }$, and which satisfy the same equation, $%
\square \xi _{\mu \nu }=0$. This means that we can choose the functions $\xi
_{\mu }$ so as to impose four conditions on $\bar{h}_{\mu \nu }$. In
particular, we can choose $\xi ^{0}$ such that the trace $\bar{h}=0$. Note
that if $\bar{h}=0$, then $\bar{h}_{\mu \nu }=h_{\mu \nu }$. The three
functions $\xi ^{i}(x)$ are now chosen so that $h^{0i}(x)=0$. Since $\bar{h}%
_{\mu \nu }=h_{\mu \nu }$, the Lorentz condition (36) with $\mu =0$ reads 
\begin{equation}
\partial ^{0}h_{00}+\partial ^{i}h_{0i}=0  \tag{44}
\end{equation}%
Having fixed $h_{0i}=0$, this simplifies to 
\begin{equation}
\partial ^{0}h_{00}=0  \tag{45}
\end{equation}%
so $h_{00}$ becomes automatically constant in time. A time-independent term $%
h_{00}$ corresponds to the static part of the gravitational interaction,
that is, to the Newtonian potential of the source which generated the
gravitational wave. The gravitational wave itself is the time-dependent part
and therefore, as far as the gravitational wave is concerned, $\partial
^{0}h_{00}=0$ means that $h_{00}=0$. So, we have set all four components $%
h_{0\mu }=0$ and we are left only with the spatial components $h_{ij}$, for
which the Lorentz gauge condition now reads $\partial ^{j}h_{ij}=0$, and the
condition of vanishing trace becomes $h_{i}^{i}=0$. In conclusion, we have
set 
\begin{equation}
h_{0\mu }=0\text{ \ \ \ \ }h_{i}^{i}=0\text{ \ \ \ \ }\partial ^{j}h_{ij}=0%
\text{ \ \ \ }  \tag{46}
\end{equation}%
This defines the transverse-traceless gauge, or $TT$ gauge [20]. By imposing
the Lorentz gauge, we have reduced the 10 degrees of freedom of the
symmetric matrix $h_{\mu \nu }$ to six degrees of freedom, and the residual
gauge freedom, associated to the four functions $\xi ^{\mu }$ that satisfy
equation (43), has further reduced these to just two physical degrees of
freedom. This is the same number of physical degrees of freedom in an
electromagnetic wave. We will denote the metric in the $TT$ gauge by $%
h_{ij}^{TT}$.

Equation (39) has the plane wave solutions 
\begin{equation}
h_{ij}^{TT}=e_{ij}(\vec{k})e^{ikx}  \tag{47}
\end{equation}%
with $k^{\mu }=\left( \omega /c,\vec{k}\right) $. The tensor $e_{ij}(\vec{k}%
) $ is called the polarization tensor. We follow the usual convention that
the real part is taken at the end of the computation. For a gravitational
wave propagating along the $z$ axis we have 
\begin{equation}
h_{ij}^{TT}(t,z)=\left( 
\begin{array}{lll}
h_{+} & h_{\times } & 0 \\ 
h_{\times } & -h_{+} & 0 \\ 
0 & 0 & 0%
\end{array}%
\right) _{ij}\cos [\omega \left( t-\frac{z}{c}\right) ]  \tag{48}
\end{equation}%
where $h_{+}$ and $h_{\times }$ are called the amplitudes of the \textsl{plus%
} and \textsl{cross} polarization of the wave.\ 

\section{Black holes, gravitational waves and quantum gravity}

In this section we present the basic ideas for a future, mathematically more
rigorous, attempt to combine black holes and gravitational waves using the
quantization of geometric quantities introduced by LQG.

\subsection{Energy spectrum for gravitational waves}

As mentioned in the introduction, LQG is a theory that attempts to describe
the quantum mechanics of the gravitational field based on the canonical
quantization of GR. From the point of view of LQG, GR is a constrained
Hamiltonian system with first-class constraints [14]. These first-class
constraints generate the gauge symmetries of GR in the Hamiltonian
formalism. The most interesting result of LQG is that geometric quantities,
such as area and volume, are quantized in terms of the Planck length $L_{P}$%
. However, the equations giving the eigenvalues of the area and volume
operators in LQG depend on the choice of an arbitrary parameter $\gamma $,
the Immirzi [16] parameter, which fixes the precise scale of the quantum
theory.

In ref. [9], using the value 
\begin{equation}
\gamma =\left( 4\pi \sqrt{3}\right) ^{-1}  \tag{49}
\end{equation}%
it was pointed out that a plane gravitational wave propagating in the
quantum space of LQG can only have wavelengths that are integer multiples of
the Planck length $L_{P}$. Since $L_{P}$ is the distance that light travels
during the Planck time $T_{P}$, the period of the gravitational wave must be
an integer multiple of the Planck time $T_{P}$. Assuming the validity of the
above conditions and using the similarities between electromagnetic and
gravitational waves, which are: a) both types of wave describe the same
number of physical degrees of freedom, b) both types of wave are transverse
waves and c) both types of wave propagate at the speed of light, ref. [9]
introduced an energy spectrum for gravitational waves. This energy spectrum
is given by 
\begin{equation}
E_{n}=h\nu _{n}=\text{%
h{\hskip-.2em}\llap{\protect\rule[1.1ex]{.325em}{.1ex}}{\hskip.2em}%
}\omega _{n}=\frac{2\pi }{n}c^{2}M_{P}\text{ \ \ \ }n=1,2,3,...\text{\ \ \ \ 
}  \tag{50}
\end{equation}%
where $M_{P}$ is the Planck mass given in equation (3) and the integer $n$
determines the energy and the number of Planck lengths contained in the
wavelength of the gravitational wave. In equation (50) $c^{2}M_{P}=1,22%
\times 10^{19}$Gev so gravitational waves with wavelengths containing a
small number of Planck lengths can have large energies. As $n\rightarrow
\infty $ the energy difference between two consecutive energy levels becomes
infinitesimal. Therefore in the $n\rightarrow \infty $ limit the energy
spectrum becomes effectively continuous, as is the case in Hawking%
\'{}%
s thermal radiation.

\subsection{Distant observers}

In the classical theory a black hole with vanishing charge and vanishing
angular momentum evolves rapidly towards the Schwarzschild solution, by
radiating away all excess energy. In the quantum theory, however, the
Heisenberg uncertainty relations prevent the black hole from converging
exactly to a Schwarzschild metric, and quantum fluctuations may remain [10].
From the point of view of this paper, the logical procedure now would be to
describe the quantum mechanical processes of emission and absorption of
gravitational radiation by quantum fluctuations of the event horizon.
However, at present, a mathematical description of such quantum mechanical
processes is still under study. In addition to the technical difficulties,
there are also conceptual difficulties. But, as we shall see in the next
subsection, the basic ideas behind the physics are very simple. This
simplicity of the basic ideas is the motivation for this paper. We will use
this subsection to consider two of the conceptual difficulties.

The Schwarzschild spacetime becomes flat at large distances from the
gravitational source and the coordinates $\left( t,r,\theta ,\varphi \right) 
$ provide a global reference frame only for an observer at infinity.
However, physical quantities measured by arbitrary observers are not
specified directly by the coordinates but rather must be computed from the
metric. Therefore, to measure a time interval for a stationary clock at $r$
we set $dr=d\theta =d\varphi =0$ in the line element (13) and we use $%
ds^{2}=-c^{2}d\tau ^{2}$ to obtain 
\begin{equation}
d\tau =\sqrt{\left( 1-\frac{2GM}{c^{2}r}\right) }dt  \tag{51}
\end{equation}%
In the above equation $d\tau $ is the \textsl{proper time} and $dt$ is the 
\textsl{coordinate time}. The physical time interval measured by a local
observer is given by the proper time $d\tau $, not by the coordinate time $%
dt $. Proper time and coordinate time will be approximately equal only if
the effect of the gravitational field is very weak. In the context of this
paper, this means that the $t$ coordinate that appears in the Schwarzschild
metric (13) can be identified with the $t$ coordinate that appears in the
gravitational wave (48) only in the reference frame of a distant observer.

Setting $dt=d\theta =d\varphi =0$ in the line element (13) gives an interval
of radial distance 
\begin{equation}
ds=\frac{dr}{\sqrt{\left( 1-\frac{2GM}{c^{2}r}\right) }}  \tag{52}
\end{equation}%
$ds$ is the \textsl{proper distance} and $dr$ is the \textsl{coordinate
distance}. The physical radial interval measured by a local observer is the
proper distance $ds$, not the coordinate distance $dr$. The proper distance
and coordinate distance will be approximately equal only for distant
observers. In addition to this, equation (52) is valid only in an arbitrary
fixed direction since, to arrive at it, we required that $d\theta =d\varphi
=0$. But the gravitational wave (48) propagates in an arbitrary $z$
direction. A partial solution for this problem comes from the fact that the
linearized version of GR is invariant under global Lorentz transformations
[20]. Therefore, for distant observers, before fixing the Lorentz gauge
condition (36) we can perform a Lorentz rotation%
\begin{equation}
x^{\mu }\rightarrow \Lambda _{\nu }^{\mu }x^{\nu }  \tag{53}
\end{equation}%
and make the $z$ direction of the distant observer%
\'{}%
s coordinate system coincident with the particular direction of the
Schwarzschild $r$ coordinate. Rotations never spoil the condition $\mid
h_{\mu \nu }\mid $ \ $\ll $ \ 1. The distant observer will then interpret
the gravitational wave as coming directly from the black hole.

\subsection{The perspective of quantum gravity}

In this subsection we use the reference frame of a distant observer and
impose a quantization process on the gravitational wave (48) and on the the
event horizon of the Schwarzschild black hole. We will find that the
concepts of a quantum gravitational wave and a quantum event horizon can be
related to each other using the concept of a quantum of length.

First we recall the usual relations $k=2\pi /\lambda $ and $\lambda \nu =c$
for a plane gravitational wave. Then we write 
\begin{equation}
\omega \left( t-\frac{z}{c}\right) =\omega t-\frac{\omega }{c}z  \tag{54}
\end{equation}%
Recalling that the angular frequency of a wave is defined as $\omega =2\pi
\nu $ we can write 
\begin{equation*}
\lambda \nu =\lambda \frac{\omega }{2\pi }=c\text{ \ }\rightarrow \text{ \ }%
\frac{\omega }{c}=\frac{2\pi }{\lambda }=k
\end{equation*}%
and therefore equation (54) becomes 
\begin{equation}
\omega \left( t-\frac{z}{c}\right) =\omega t-kz  \tag{55}
\end{equation}%
From the energy spectrum (50) we quantize the angular frequency as%
\begin{equation}
\omega _{n}=\frac{E_{n}}{\text{%
h{\hskip-.2em}\llap{\protect\rule[1.1ex]{.325em}{.1ex}}{\hskip.2em}%
}}  \tag{56}
\end{equation}%
and introduce the quantized wave number as $k_{n}=2\pi /\lambda _{n}$ with $%
\lambda _{n}$ giving the number of Planck lengths $L_{P}$ contained in the
wavelength of the gravitational wave [9]. Finally we use de Broglie%
\'{}%
s relation $p=h/\lambda $ and introduce the quantized momentum variable 
\begin{equation}
P_{n}=\frac{h}{\lambda _{n}}=\frac{h}{2\pi }\times \frac{2\pi }{\lambda _{n}}%
=\text{%
h{\hskip-.2em}\llap{\protect\rule[1.1ex]{.325em}{.1ex}}{\hskip.2em}%
}k_{n}  \tag{57}
\end{equation}%
Using the above equations, a quantized gravitational wave propagating in the 
$z$ direction, in the transverse-traceless gauge can be written as 
\begin{equation}
h_{ij}^{n}(t,z)=\left[ 
\begin{array}{lll}
h_{+} & h_{\times } & 0 \\ 
h_{\times } & -h_{+} & 0 \\ 
0 & 0 & 0%
\end{array}%
\right] _{ij}\cos [\frac{1}{\text{%
h{\hskip-.2em}\llap{\protect\rule[1.1ex]{.325em}{.1ex}}{\hskip.2em}%
}}\left( E_{n}t-P_{n}z\right) ]\text{ \ \ \ \ \ }n=1,2,3,...  \tag{58}
\end{equation}%
where $n$ gives the possible energy values, the possible momentum values and
the number of Planck lengths $L_{P}$ contained in the wavelength of the
wave. We can use the quantum gravitational wave (58) as a basis for the
interpretation of $L_{P}$ as a quantum of length.

Using the idea that $L_{P}$ defines a quantum of length we now quantize the
Schwarzschild black hole event horizon area using the fact that the event
horizon is spherically symmetric. We start by noting that the area of the
horizon is $A=4\pi R_{S}^{2}$ and the circumference of the horizon is $%
L=2\pi R_{S}$. Therefore the area of the event horizon is 
\begin{equation}
A=\frac{L^{2}}{\pi }  \tag{59}
\end{equation}%
We now impose the quantum condition that the circumference $L$ of the event
horizon must contain an integer number $N$ of Planck lengths $L_{P}$, that
is $L=NL_{P}.$ Inserting this condition into equation (59) we obtain the
quantized area of the event horizon 
\begin{equation}
A=\frac{1}{\pi }N^{2}L_{P}^{2}  \tag{60}
\end{equation}%
Inserting equation (60) into equation (16) for the entropy of a classical
black hole, we obtain the entropy for a black hole with a quantized event
horizon area 
\begin{equation*}
S=\frac{c^{3}k_{B}}{4\text{%
h{\hskip-.2em}\llap{\protect\rule[1.1ex]{.325em}{.1ex}}{\hskip.2em}%
G}}A
\end{equation*}%
\begin{equation*}
\text{ \ \ \ }=\frac{k_{B}}{4L_{P}^{2}}A
\end{equation*}%
\begin{equation*}
\text{ \ \ \ \ \ \ \ \ \ \ \ \ }=\frac{k_{B}}{4L_{P}^{2}}\frac{1}{\pi }%
N^{2}L_{P}^{2}
\end{equation*}%
\begin{equation}
\text{ \ \ \ \ \ \ \ \ }=\frac{k_{B}}{4\pi }N^{2}\text{ \ \ \ \ }  \tag{61}
\end{equation}%
From equation (61) we see that what gives rise to the Schwarzschild black
hole entropy (16) is the number of quantum of length $L_{P}$ contained in
the circumference of the event horizon.

Let us now consider the black hole mass $M$ given by equation (14).
Combining equations (14) and (60) we obtain for the mass of the black hole%
\begin{equation*}
M=\sqrt{\frac{c^{3}A}{16\pi G^{2}}}
\end{equation*}%
\begin{equation*}
=\sqrt{\frac{c^{3}N^{2}L_{P}^{2}}{16\pi ^{2}G^{2}}}
\end{equation*}%
\begin{equation*}
\text{ \ \ \ \ }=\frac{N}{4\pi \sqrt{c}}\sqrt{\frac{\text{%
h{\hskip-.2em}\llap{\protect\rule[1.1ex]{.325em}{.1ex}}{\hskip.2em}%
c}}{G}}
\end{equation*}%
\begin{equation}
\text{ \ }=\frac{N}{4\pi \sqrt{c}}M_{P}  \tag{62}
\end{equation}%
\ Equation (62) shows that the mass of the black hole is quantized in terms
of the Planck mass $M_{P}$, given in equation (3). This result confirms the
consistency of the ideas we present in this paper. Combining now equation
(62) with equation (4) we have 
\begin{equation}
M=\frac{\sqrt{c^{3}}}{4\pi G}NL_{P}  \tag{63}
\end{equation}%
Equation (63) shows that a black hole can increase its mass to a mass 
\begin{equation}
M=\frac{\sqrt{c^{3}}}{4\pi G}\left( N+n\right) L_{P}  \tag{64}
\end{equation}%
by absorbing a quantum gravitational wave of the type (58) with a wavelength
containing $n$ quanta of length $L_{P}.$ In the opposite process, a black
hole can decrease its mass to a mass 
\begin{equation}
M=\frac{\sqrt{c^{3}}}{4\pi G}\left( N-n\right) L_{P}  \tag{65}
\end{equation}%
\ by emitting a quantum gravitational wave of the type (58) with a
wavelength containing $n$ quanta of length $L_{P}$. From equation (65) we
see that a black hole can completely convert its mass into quantum
gravitational radiation by emitting $N$ quanta of length $L_{P}$. Notice
that in this case there is no information loss paradox. The mass of the
black hole is converted in quantum gravitational radiation with well-defined
polarization, energy and momentum.

\subsection{Conclusions}

The objective of this paper is to expose the basic ideas that will give
support to a future, mathematically more rigorous, study of the emission of
quantum gravitational radiation by black holes. For this purpose, in section
two we reviewed the Schwarzschild vacuum solution for the Einstein equations
of GR. This solution describes a spherically symmetric spacetime with a
black hole and an event horizon. We displayed the equation that relates the
black hole mass to the area of the event horizon, the equation that gives
the black hole temperature due to Hawking%
\'{}%
s thermal radiation and the equation for the black hole entropy. We also
briefly described how the equation for the black hole entropy emerges in the
framework of LQG and why the degrees of freedom responsible for the entropy
remain undetermined in this framework.

In section three we considered the propagation of gravitational waves. We
reviewed the process of linearization of the Einstein equations. We
described in detail the steps for the elimination of the unphysical degrees
of freedom of the spacetime metric to arrive at the equations of motion in
the transverse-traceless gauge. The solution for a gravitational wave
propagating along the $z$ direction was displayed.

Section four contains our contribution to the subject. We reviewed the
energy spectrum for a gravitational wave propagating in the quantized
spacetime of LQG which was obtained in [9]. We reviewed the notion of a
distant observer. Then, using the energy spectrum obtained in [9] and de
Broglie%
\'{}%
s relation, we imposed a quantization process on the classical gravitational
wave described in section three. Since the quantum gravitational wave can
only have wavelengths which are integer multiples of the Planck length $%
L_{P},$ we interpreted $L_{P}$ as defining the quantum of length. Using the
spherical symmetry of the Schwarzschild black hole and the notion of the
quantum of length we quantized the area of the event horizon and found that
the microstate responsible for the black hole entropy is the quantum of
length $L_{P}.$ Using the quantized event horizon area, we showed that the
mass of the black hole is quantized in terms of the Planck mass $M_{P}$, a
result that confirms the ideas presented in this paper. Finally, using the
relation between the Planck mass $M_{P}$ and the quantum of length $L_{P},$
we displayed equations that show that a black hole can increase or decrease
its mass by absorbing or emitting quantum gravitational radiation. In
particular, the black hole can convert all its mass into quantum
gravitational radiation with well defined polarization, energy and momentum,
thus avoiding the information loss paradox.

\end{document}